# Dielectric secondary relaxation of water in aqueous binary glass-formers

Johan Sjöström,[a] Johan Mattsson,[b] Rikard Bergman, Erik Johansson, Karin Josefsson, David Svantesson, and Jan Swenson

The dielectric relaxation of water in glassy aqueous binary mixtures exhibits an Arrhenius behaviour with a nearly universal activation energy. We here demonstrate that its characteristic relaxation time follows a remarkably general functional dependence on the weight fraction of water for a wide range of molecular systems.

## 1. Introduction

Water is one of the most important substances on earth, playing essential roles in geology, in chemistry and in most biological systems[1]. The relaxation dynamics of hydration water in proteins, for instance, is believed to control important dynamics of the proteins themselves and thus to influence the process of protein folding and functionality[2,3].

Surprisingly, many of the physical properties of water are not well understood and its dynamics, structure and thermodynamics display 'anomalous' behaviour compared to most other liquids; this is believed to be a direct consequence of the strong hydrogen (H-) bonding that characterizes the intermolecular interactions in water[4].

A problem when investigating pure water and a contributing reason why water is still not well understood is its strong tendency to crystallize. As a consequence, water is very difficult to study in its supercooled regime, where many of its apparent anomalies are located. Below the homogeneous nucleation temperature at ~231 K, one has to resort to indirect methods to learn about its behaviour and most studies have been performed either on water in confinement[5-8] or on water mixed with other liquids[9-14].

A liquid in its supercooled state generally displays several molecular relaxation processes. The primary $\alpha$ relaxation is related to the viscosity of the liquid and slows down dramatically as the temperature is reduced, which eventually causes the system to fall out of equilibrium at the glass transition temperature, $T_g$. In addition, a faster secondary so called $\beta$ relaxation is generally observed either as a distinct relaxation process or as a high frequency contribution to the $\alpha$ relaxation, a so called excess wing[11].

Moreover, a range of studies[12-15] have shown that when water is mixed with another glass-forming liquid a new dielectric relaxation process appears. This additional relaxation process shows a number of general features: (i) it follows an Arrhenius temperature dependence in the glassy state, $\tau = \tau_0 \exp(E/k_B T)$, where $E$ is the activation energy and $\tau_0$ the relaxation time in the high temperature limit (ii) the shape of its response function is symmetric, or very near symmetric, on a logarithmic frequency scale (iii) $E$ has a value ~0.54 eV (Ref. 14,16), independent of the system in which water is dissolved or the amount of water that is dissolved[14] and (iv) its relaxation strength increases systematically with increasing water content.

We note that (i)-(ii) are characteristics typical of secondary relaxations in glass-forming systems and the value of the activation energy in (iii) suggests a relatively 'local' character of the observed response; $E$ roughly corresponds to twice the energy of a typical H-bond[17]. Furthermore, it is interesting to note that a secondary relaxation with similar properties (i)-(iii) is observed also for pure water studied in hard confinement, where crystallization can be suppressed[8]. This together with the behaviour of its strength (iv), further suggests that the relaxation observed in aqueous mixtures is largely characteristic of the behaviour of water itself. Interestingly, the secondary relaxation observed for water in confinement has been assigned as the $\beta$ relaxation of pure water, corresponding to the $\beta$ relaxation observed for most pure glass-formers[18-22].

One difference between the secondary relaxation observed in aqueous mixtures and most "normal" $\beta$ relaxations is the quantitative values of its Arrhenius prefactor, $\tau_0$. Typically, when fitting relaxation data for aqueous mixtures in their glassy state to an Arrhenius expression, values of $\tau_0 \sim 10^{-22}$-$10^{-16}$ s are found, which are clearly different from characteristic molecular vibrational time-scales, $\tau_0 \sim 10^{-14}$-$10^{-12}$ s. This surprising behaviour demonstrates our present lack of understanding of water in general and of water mixed with other liquids in particular and stresses the need for systematic experimental investigations.

In this study we investigate the concentration dependence of the water-related secondary relaxation, here denoted the $w$ relaxation, for binary mixtures of water and a series of liquids for which the H-bond density can be systematically varied[22].

## 2. Experimental

We study oligomeric liquids based on the same monomeric unit, propylene oxide, but varying in chain-ends: X-[$CH_2CH(CH_3)O$]$_n$-H, where X=HO for propylene glycols and X=$CH_3$-O for propylene glycol monomethyl ethers. We investigate the monomers with $n=1$ (PG, PGME), dimers with $n=2$ (2PG, 2PGME) and trimers with $n=3$ (3PG, 3PGME), respectively. For both the glycols and the monomethyl ethers, H-bonding plays an important role[22], but as shown in Fig. 1, and discussed below, an increasing water content has a completely different effect on the $T_g$-values and thus the $\alpha$ relaxations, in the two types of liquids.

All samples were purchased from Sigma-Aldrich. They were freeze dried and their purity was confirmed by IR-spectroscopy. The liquids were mixed with distilled milli-Q water (conductivity < 0.1 µS/m) followed by homogenisation for fifteen minutes in an ultrasonic bath.

The differential scanning calorimetry (DSC) experiments were performed on a TA Instrument (DSC Q1000). The samples were initially cooled into their glassy states from room temperature at a rate of -30 K/min. Subsequently, the $T_g$ values were determined as the onset of the heat capacity step upon heating from the glass at a rate of +10 K/min.

No sign of crystallisation was observed during the initial cooling for mixtures with water concentrations $C_w$ < 60 wt % (mass of water to total mass of sample) with the exception of the PGME mixtures which showed crystallization at $C_w$ > 54 wt %. For the latter samples, crystallization was avoided by quenching of the samples into a bath of liquid nitrogen followed by direct transfer to a pre-cooled cryostat; for these samples crystallization could not be directly monitored on cooling. However, all samples, both quenched and unquenched, showed one distinct glass transition.

The dielectric response was measured using a Novocontrol Alpha Analyser in the frequency range 14 mHz - 3 MHz. The 100 micron thick samples, for which the thickness was controlled by silica spacers, were sandwiched between electrodes with a diameter of 20 mm placed within a sealed sample cell. Generally, for all six systems the samples were cooled from room temperature using controlled cryostat cooling at a typical rate of -20 K/min; no sign of crystallinity was found for T<205 K using this protocol. The two PGME samples with the highest water fraction were instead quenched by submerging the sample cell into liquid nitrogen followed by placing the sample cell into a pre-cooled cryostat.

## 3. Results and discussion

As water is added to the pure $n$PG liquids, the $T_g$ value decreases but reaches a plateau at ~20 wt %; the plateau behaviour continues to ~40-50 wt %, where an onset to a stronger $T$ dependence takes place leading to a decrease of $T_g$ upon further addition of water. In contrast, for aqueous $n$PGME $T_g$ increases strongly upon addition of water and reaches a maximum for $C_w$ =45 wt % (55 wt % for PGME). For higher $C_w$, the $T_g$ values decrease rapidly, as shown in Fig. 1. The observed behaviour for the monomethyl ethers with an increasing $T_g$ value for increasing water content is a highly unusual behaviour for aqueous systems. We investigate this behaviour in detail in a separate publication[23], but we suggest that the increase of $T_g$ for low to moderate $C_w$ is an effect of the build-up of H-bond mediated structures and the decrease of $T_g$ for the highest $C_w$ is likely due to a plasticizing effect of water saturated structures.

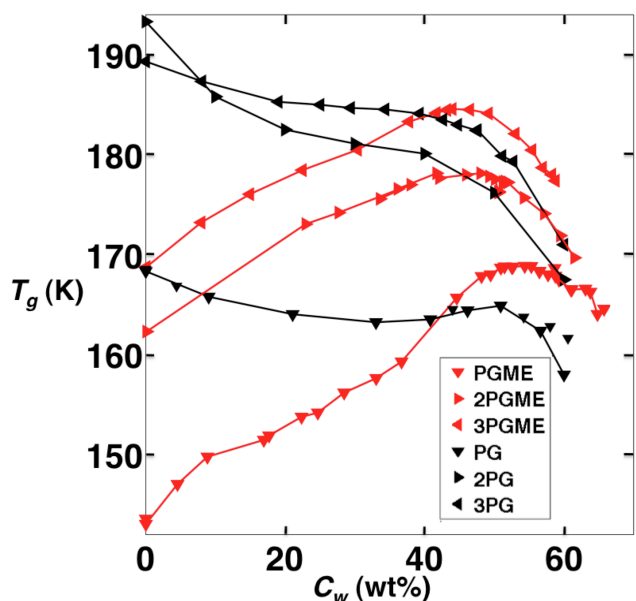

**Fig. 1:** Calorimetric glass transition temperatures for aqueous mixtures of $n$PG and $n$PGME, respectively, as a function of water content. The solid lines are guides to the eye.

In the following, we focus primarily on the behaviour of the $w$ relaxation and for this study it is sufficient to recognize the marked difference between the $T_g$ behaviours of the glycols and the monomethyl ethers.

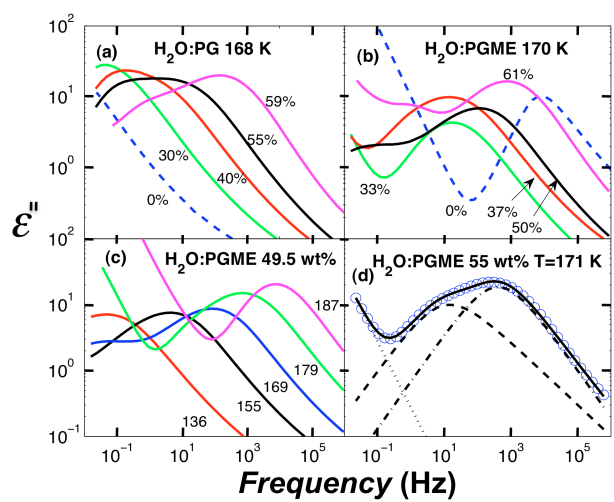

**Fig. 2:** Dielectric loss spectra. Upper left: Water-PG mixtures with different water concentrations for $T$=168 K. Upper right: Water-PGME mixtures for $T$=170 K. Lower left: Water-PG mixtures at different temperatures (in units of K) for a $C_w$ =49.5 wt %. Lower right: Example of fitting result. (··): power law dc-conductivity; (--): HN-function, $\alpha$ relaxation; (-··-): CC function, w-relaxation; (solid line): total fit.

The dielectric loss, $\varepsilon''$, is shown for aqueous PG and PGME in Fig. 2. In addition to the $\alpha$ relaxation, PGME exhibits a secondary $\beta$ relaxation loss peak at frequencies above the $\alpha$ loss. Also PG shows a secondary response, but this relaxation is submerged under the high frequency flank of the $\alpha$ relaxation and is thus only observed as an excess wing. As water is added, for both PG and PGME, a third relaxation process, the $w$-relaxation, is observed in between the $\alpha$ and the β relaxations (see Supplementary Information for an example of the spectra at low water contents). The strength of this additional process increases monotonically with water content and even for modest water concentrations the contribution of the β relaxation to $\varepsilon''$ is negligible. This behaviour makes an analysis of the β relaxations of the mixtures difficult and we therefore restrict our discussion to the α relaxation and the additional w-relaxation. Analogous behaviours of the α, the β and the additional w-relaxation are observed for both the dimers and trimers of PG and PGME.

We note that for the PGME mixtures with water concentrations above the maximum in $T_g$, we found evidence for an additional relaxation process. This process is several order of magnitude slower than the w relaxation, but has roughly the same activation energy. However, due to the difficulty in resolving, and thus quantitatively describing, this relaxation process, it will not be discussed further in the following.

The dielectric response of the aqueous mixtures can be well described with a sum of standard fit functions; a Havriliak-Negami[24] (HN) function for the α relaxation and a Cole-Cole[25] (CC) function for the w relaxation (Supplementary Information). For anhydrous samples or samples with low water content, for which the β relaxation can be resolved, it is described by a CC function. The DC-conductivity is well described by a power-law term with an exponent of -1 in the dielectric loss.

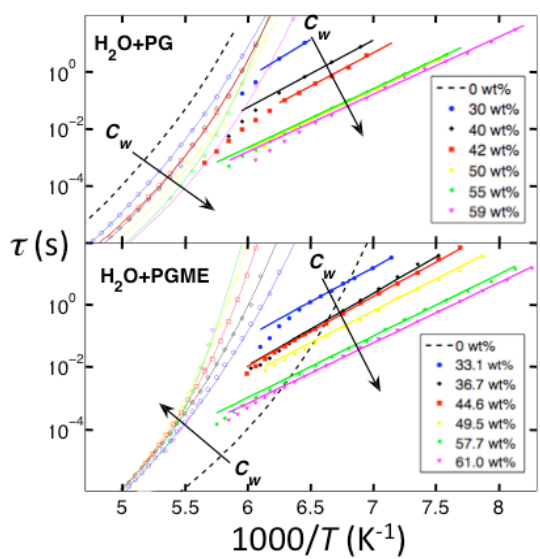

**Fig. 3:** Dielectric relaxation times of the water-PG mixtures (top) and water PGME mixtures (bottom). The dashed black lines are the α relaxation of anhydrous samples. Solid symbols are relaxation times of the w relaxation and thick lines are the fits to the Arrhenius equation. Open symbols are the relaxation times of the α relaxation and thin lines the fits to the VFT equation. The arrows indicate the trend of increasing water content.

From our fits, shown in Fig. 2, we extract the characteristic relaxation times for the α and w relaxations. The resulting relaxation times for PG and PGME at different hydration levels are shown in an Arrhenius representation in Fig. 3 (the corresponding plots for the dimers and trimers are included in the Supplementary Information). The temperature dependent α relaxation times for all investigated samples are well described by Vogel-Fulcher-Tamman[26] (VFT) functions, as expected for α relaxations and demonstrated by the fits in Fig. 3. For temperatures near $T_g$, the α relaxation generally becomes faster as water is added to the glycols, as shown in Fig. 3. In contrast, the opposite behaviour with a slowing down of the α relaxation for increasing water content is observed for monomethyl ethers up to a certain value of $C_w$, above which τ decreases. We find that the dielectric $T_g$ values, determined as the temperatures where $\tau_\alpha$=100 s, correspond well to the calorimetric $T_g$ values.

The characteristic relaxation times for the w relaxations exhibit Arrhenius temperature dependences in the glassy state for all samples. This behaviour changes above $T_g$, where the VFT behaviour of the α relaxation is approached; this is commonly observed for secondary relaxations in supercooled liquids[27] and will not be further discussed here. Here we focus on the behaviour of the w relaxation within the glassy state. Remarkably, in marked contrast to the very different behaviours of the α relaxations for PG and PGME, the w relaxation shifts systematically towards shorter times as more water is added for both systems. We find the same behaviours also for the dimers and trimers, as shown in the Supplementary Information.

The activation energy for the w relaxation is highly similar for all investigated systems with only a weak decrease observed for increasing $C_w$; values within the range $E$=0.46±0.06 eV (44±6 kJ mol$^{-1}$.) were found for all mixtures, which is consistent with the previously reported values for available literature data[14].

To investigate this behaviour further we plot the w relaxation time at a fixed temperature, $T$=150 K, as a function of $C_w$ for all six systems in Fig. 4; at this temperature all samples are in their glassy states. Surprisingly, we find that all systems show the same quantitative exponential decrease of $\tau_w$(150 K) with $C_w$. We recall the markedly different $C_w$ dependence of the α relaxations for the different liquids and we thus establish that for these systems the w relaxation is essentially insensitive both to the route of glass formation and to the detailed nature of its glassy environment.

Even more surprising is the fact that the observed correlation holds for an even wider range of different liquids. In the inset to Fig. 4, we have included all, to our knowledge, available literature data for the w relaxation in aqueous binary systems that are glassy at 150 K. The quantitative agreement between the different data sets is surprising, particularly since we expect the spatial distribution of water to be different for different systems. Decreased interactions between water and the matrix molecules due to the formation of water clusters at higher $C_w$ could lead to faster dynamics, but such an effect should depend on the density of water clusters and not on the overall water content, $C_w$.

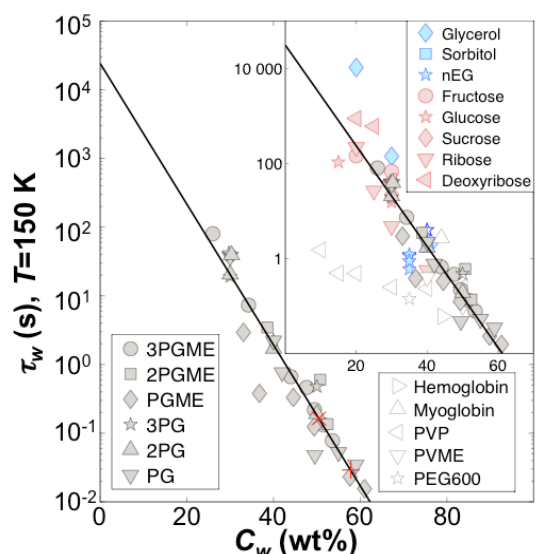

**Fig. 4** Dielectric relaxation time of the $w$-relaxation, $\tau_w$, at 150 K for all systems in this study against $C_w$. The inset shows $\tau_w$ for literature data on a wide range of different systems including glycerol[28], sorbitol, nEG[14], fructose and sucrose[29], glucose[30], Ribose and deoxyribose[31], Hemoglobin and myoglobin[32], PVP[33], PVME and PEG600[14]. Solid lines are exponential fits to the data excluding polymeric systems (green symbols). Relaxation time for pure ware confined in clay[7] (x) and MCM-41[8] (+) are also shown.

However, as seen from the inset to Fig. 4, some polymeric and biopolymeric systems do display a clearly different behaviour at least at low (<35 wt%) hydration levels. This might be due to a highly uneven distribution of water at lower concentrations, which would lead to local water concentrations significantly different from the average values reported in Fig. 4, making direct comparisons difficult. The fact remains that the observed correlation is remarkably general and excluding the polymeric systems from the discussion, we can describe all systems by the same exponential behaviour, as demonstrated by the solid line in Fig. 4; this behaviour generally correlates overall water concentrations (where no crystallization occur) and $\tau_w$ within an accuracy of a decade in time. Thus, the main cause of the change in $\tau_w$ is a decrease of the concentration dependent 'attempt' time scale, $\tau_0(C_w) = \tau' \exp(-C_w/C_w^*)$, where $\tau'$ is the time scale for $C_w=0$ and $C_w^*$ characterizes the speeding up of $\tau_0$ with increasing $C_w$.

## 4. Concluding remarks

We have thus found that the secondary dielectric relaxation of water in glassy binary mixtures: (i) follows an activated Arrhenius temperature dependence with an almost system independent activation energy and (ii) its relaxation time at a fixed temperature decreases exponentially with increasing weight fraction of water.

An extrapolation of the exponential behaviour in Fig. 4 to $C_w=0$ gives $\tau'=4\cdot10^{-14}$ s, consistent with a typical vibrational time-scale expected for a pure glass-former and we find that $C_w^* \approx 10$ wt %. Moreover, it is interesting to note that for the highest investigated water concentrations we obtain a behaviour corresponding well to that observed for pure water measured in hard confinement, where crystallization can be avoided[7,8]. To demonstrate this we have included the values of $\tau_w(150\ K)$ determined from dielectric measurements on water within two different confinements[7,8] in Fig. 4.

It is interesting to note that at the highest $C_w$ values where the data coincide, a marked change in behaviour is observed in the $T_g$ vs. $C_w$ dependence, observed either as a maximum (monomethyl ethers) or as a rapid decrease (glycols). This suggests that these characteristic behaviours, observed for the $\alpha$ relaxation, are related to the onset of some degree of 'bulk-like' water properties in the mixtures and suggests a possible saturation of the $w$ relaxation time for higher concentrations if crystallisation could be prevented.

In summary, we demonstrate that the unique secondary relaxation observed in binary aqueous mixtures exhibits a remarkably general dependence on the overall water weight fraction. We determine a quantitative relationship, which describes this dependence for a wide range of aqueous mixtures. We find that this near 'universal' effect is unrelated to the highly system dependent effects of water content on the structural relaxation time. The importance of better understanding the role played by 'local' water dynamics in biology and in many industrial processes, such as the cryopreservation of biomolecules, makes further investigations of the generalities presented in this work an important avenue for future work.

This work was financially supported by the Swedish Research Council, the Swedish Foundation for Strategic Research and the Knut and Alice Wallenberg Foundation (JM).

## Notes and references

*Department of Applied Physics, Chalmers University of Technology, SE-41296 Göteborg, Sweden.*
*[a] E-mail: johan.sjostrom@chalmers.se*
*[b] E-mail: johanm@chalmers.se*
† Electronic Supplementary Information (ESI) available: Temperature dependent dielectric relaxation times of aqueous 2PG, 3PG, 2PGME and 3PGME. See DOI: 10.1039/b000000x/